\documentclass[pra,twocolumn,superscriptaddress]{revtex4-2}

\usepackage{nicefrac,amsmath,amssymb,mathrsfs}
\usepackage{graphicx}
\usepackage{natbib}
\usepackage{bm,color}

\begin{document}

\title{Theoretical description of optofluidic force induction}

\author{Marko \v{S}imi{\'c}}
\affiliation{Brave Analytics GmbH, Austria}
\affiliation{Gottfried Schatz Research Center, Division of Biophysics, Medical University of Graz, Neue Stiftingtalstra\ss e 2, 8010 Graz, Austria}
\affiliation{Institute of Physics, University of Graz, Universit\"atsplatz 5, 8010 Graz, Austria}

\author{Christian Hill}
\affiliation{Brave Analytics GmbH, Austria}
\affiliation{Gottfried Schatz Research Center, Division of Biophysics, Medical University of Graz, Neue Stiftingtalstra\ss e 2, 8010 Graz, Austria}

\author{Ulrich Hohenester}
\affiliation{Institute of Physics, University of Graz, Universit\"atsplatz 5, 8010 Graz, Austria}

\date{\today}

\begin{abstract}
Optofluidic force induction (\textsc{of2}i) is an optical nanoparticle characterization scheme which achieves real-time optical counting with single-particle sensitivity and high throughput.  In a recent paper [\v{S}imi{\'c} \textit{et al.}, Phys. Rev. Appl. \textbf{18}, 024056 (2022)], we have demonstrated the working principle for standardized polystrene nanoparticles, and have developed a theoretical model to analyze the experimental data.  In this paper we give a detailed account of the model ingredients including the full working equations, provide additional justification for the assumptions underlying \textsc{of2}i, and discuss directions for further developments and future research.
\end{abstract}
\maketitle

\section{Introduction}

Nanoparticle characterization in dispersion proves to be a challenging task, in particular for complex and heterogeneous particle systems~\cite{iso:16}.  Effects such as particle agglomeration and aggregation can lead to highly polydisperse or multi-modal systems, thus calling for robust, accurate and versatile characterization methods~\cite{anderson:13}.  Existing technologies, such as nanoparticle tracking analysis~\cite{hole:13} or electron microscopy, provide possibilities for single particle analysis, however, with the bottleneck of low particle throughput and offline measurements.

The recently introduced optofluidic force induction (\textsc{of2}i) technique addresses these problems using the principle of optical tweezers in combination with a continuous flow, in order to perform single particle analysis of polydisperse samples in real-time~\cite{simic:22}.  The physics underlying this scheme is similar to optical tweezer experiments, where a strongly focused laser beam is used to optically trap particles in three dimensions.  The basic principle has been pioneered by Arthur Ashkin in 1970, and has been awarded the Nobel Prize for Physics in 2018~\cite{ashkin:70}.  Optical tweezers allow for precise control of orientation, position and arrangement of the particles under investigation~\cite{lee:04,butaite:19,donato:16}.  Besides holding particles in place, a weakly focused laser beam can also achieve two-dimensional optical trapping, where the particles can move along the optical axis of the exciting beam.  Within the context of nanoparticle characterizations, this can be employed for optical chromatography~\cite{imasaka:95}.

At the heart of optical tweezers simulations lies the calculation of the optical forces~\cite{jones:15,gennerich:17}.  These forces arise from the light-matter interaction of the exciting laser beam with a particle and the resulting photon momentum transfer, ultimately leading to a light scattering problem.  While it is well established how such scattering problems can be solved for usual plane wave excitations within Mie theory~\cite{bohren:83}, more attention is required when dealing with higher-order laser modes carrying orbital angular momentum~\cite{allen:92,franke:08,shen:19}, such as the Laguerre-Gaussian beams used in \textsc{of2}i.  Again the light scattering theory for such exctations has been developed elsewhere~\cite{kiselev:14,gutierrez-cuevas:18}, but must be put together with the other ingredients of a full simulation approach with sufficient care.  Here, in addition to optical forces, viscous drag and thermal fluctuations contribute considerably to the dynamics of a particle in a liquid medium~\cite{bui:17}. 

In this paper we develop and discuss a four-step model for the simulation of \textsc{of2}i, which accounts for the incoming electromagnetic fields of a Laguerre-Gauss beam, solves Maxwell's equations for such excitations and spherical particles, computes from the solutions of Maxwell's equations the optical scattering spectra and optical forces, and uses Newton's equations of motion to simulate the particle trajectories.  A number of representative and prototypical setups are investigated to estimate the importance of the various ingredients entering our model.  

The outline of the paper is as follows. In Sec.~\ref{sec:theory} we present the theory and derivation of the OF2i trajectory model. The resulting particle trajectories are presented in Sec.~\ref{sec:results}, and we provide detailed discussions of the impact of particle refractive indices, sphere sizes, and Brownian motion.  Finally, in Sec.~\ref{sec:summary} we summarize our results and give an outlook to future work.  Some of the theoretical details are given in the Appendices.

\section{Theory}\label{sec:theory}

\begin{figure}[t]
\includegraphics[width=\columnwidth]{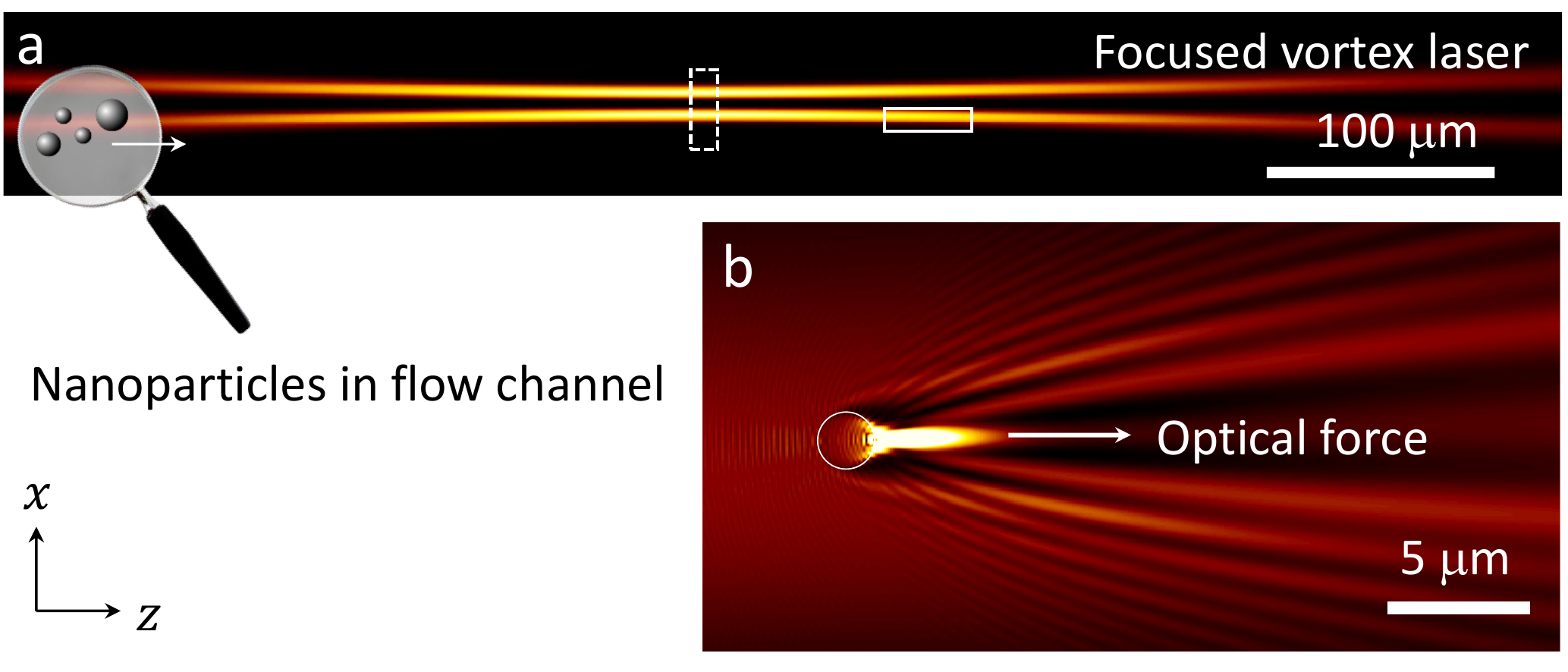}
\caption{Schematics of optofluidic force induction (\textsc{of2}i).  (a)~Nanoparticles to be analyzed are transported through a microfluidic channel alongside a weakly focused laser beam with an optical vortex (optical angular momentum $m=2$).  The dashed box indicates the region where the field distribution is shown in Fig.~\ref{fig:optforce01}.  The solid box indicates the region where in panel (b) the field intensity distribution of a nanosphere with a diameter of 2 $\mu$m, located at the intensity maximum, is shown.}
\label{fig:setup}
\end{figure}

The basic principle of \textsc{of2}i is sketched in Fig.~\ref{fig:setup}.  The nanoparticles to be analyzed are immersed in a solution and are pumped through a microfluidic flow cell.  Additionally, a weakly focused laser beam propagates in the flow direction.  The purpose of this laser is three-fold.  First, the optical forces in the transverse directions $x$,$y$ (see Fig.~\ref{fig:setup}) push the nanoparticles to the intensity maxima of the laser field, such that particles propagating sufficiently close to the maxima become trapped in the transverse directions.  Second, the optical forces in the laser propagation direction $z$ push the particles and lead to velocity changes depending on size and material properties.  Third, light is scattered off the particles and can be monitored outside the flow cell.  By analyzing the velocity changes of the individual particles being transported through the focus region, one obtains detailed information about their properties.  The light scattering intensities and emission patterns provide additional information, as will be discussed in more detail below.

An important ingredient of \textsc{of2}i is the use of a vortex laser beam with an orbital angular momentum (\textsc{oam})~\cite{allen:92,franke:08,bliokh:15,shen:19}.  Throughout this work we consider a weakly focused Laguerre-Gaussian laser beam with a polarization along $x$, with the electric field~\cite{song:20} (see also Appendix~\ref{sec:LG})
\begin{equation}\label{eq:oam}
  \bm E(r,\phi,z)\approx \mathscr{E}_m(r,z) e^{im\phi}\,\hat{\bm x}\,,
\end{equation}
where $m$ is the so-called topological charge associated with the \textsc{oam}, and $\mathscr{E}_m(r,z)$ is the field profile in the radial and propagation directions.  The intensity profile of such a beam is depicted in Fig.~\ref{fig:setup} for $m=2$.  Because of the topological charge, it has a ring-like distribution in the transverse directions with zero intensity in the center, and the trapped nanoparticles move along spiral-shaped trajectories through the focus region.  This has the advantage that nanoparticles can bypass each other more easily and collisions are strongly surpressed in comparison to laser beams with an intensity maximum on the optical axis.

In Ref.~\cite{simic:22} we have experimentally demonstrated the working principle of \textsc{of2}i for an ensemble of standard polystyrene nanoparticles with well-known size distributions, and have developed a theoretical model for the analysis of the experiments.  In the remainder of this section, we give a detailed presentation of the various ingredients entering this model.  We start by presenting the theory in its most general form, and then specialize on the implementations using either Mie theory or a fully numerical simulation approach.

\subsection{Four-step model for OF2i}\label{sec:fourstep}

\begin{figure}[t]
\includegraphics[width=\columnwidth]{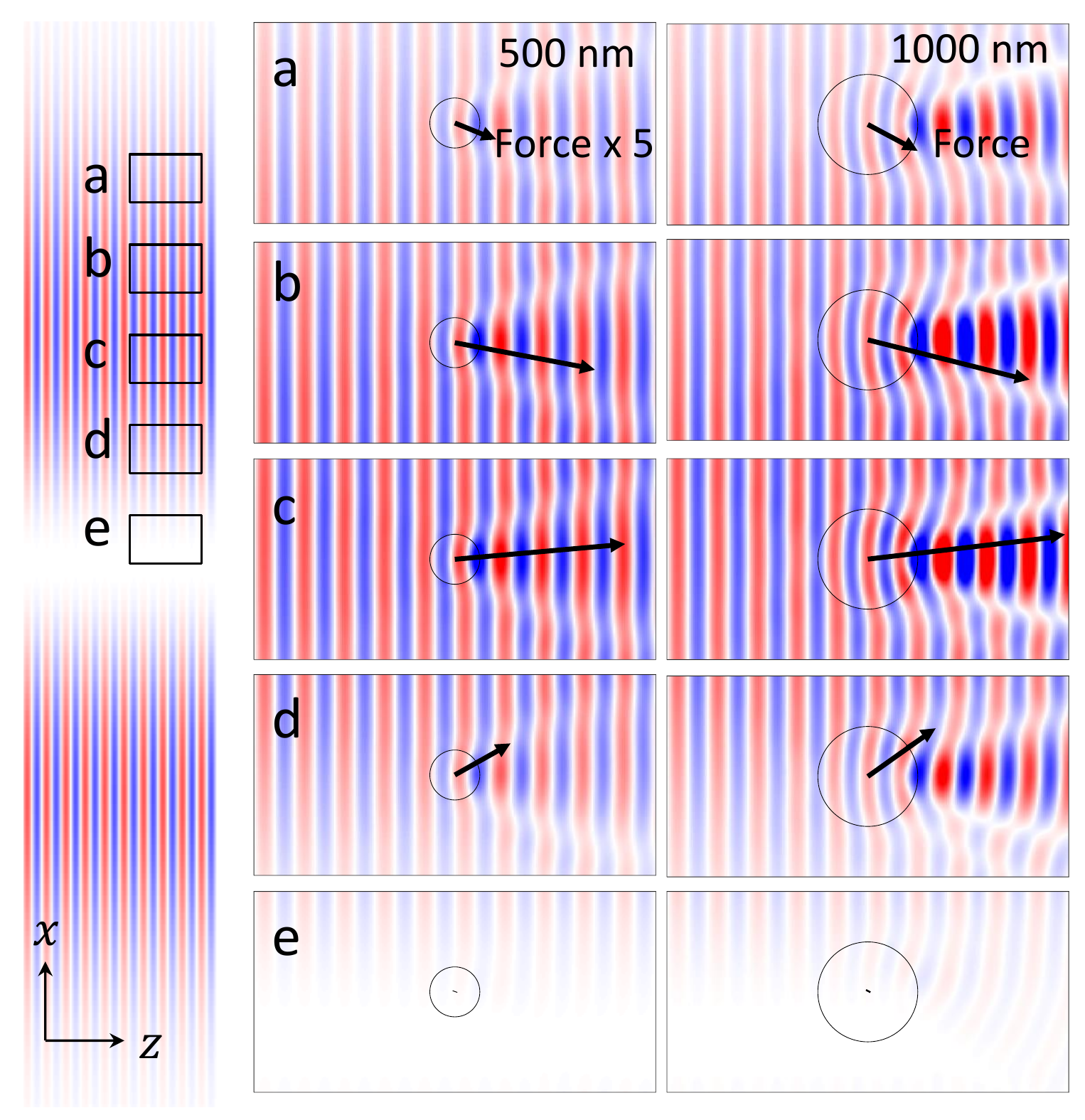}
\caption{Field distribution in the focus region of the laser, see also dashed box in Fig.~\ref{fig:setup}(a).  The light becomes deflected by the nanoparticle, through the actio-reactio principle an optical force (solid lines) is exerted on the particle that leads to a trapping in the transverse $x$,$y$ directions and a velocity change in the $z$ direction.  The positions of panels (a--e) are reported in the panel on the left.  We show results for nanospheres with diameters of 500 and 1000 nm, respectively, the refractive index is $n_b=1.33$ for the embedding medium (water) and $n=1.59$ for the nanosphere (polystyrene).  Note that in panels (e) the intensity is low and the fields are hardly visible.}
\label{fig:optforce01}
\end{figure}

The theoretical description of \textsc{of2}i consists of an electromagnetic part and a particle trajectory part.  We first provide a brief summary of the theoretical ingredients and then ponder on the details.  In the electromagnetic part, we account for the optical response of the nanoparticles and compute the optical forces and scattering fields, see also Fig.~\ref{fig:optforce01}.   We start with the incoming fields of the Laguerre-Gauss laser beam, $\bm E_{\rm inc}$, $\bm H_{\rm inc}$, which would be solutions of Maxwell's equations in absence of the nanoparticle.  In presence of the nanoparticle we additionally  have to consider the scattered fields $\bm E_{\rm sca}$, $\bm H_{\rm sca}$, which are chosen such that the boundary conditions of Maxwell's equations are fulfilled at the particle boundary.  The sum of incoming and scattered fields then provides us with the total fields, which are the proper solutions of Maxwell's equations.  From the deflection of the incoming fields we can compute the optical force $\bm F_{\rm opt}$, as shown in Fig.~\ref{fig:optforce01} and discussed in more detail below.  In the particle trajectory part, we consider a Newton's equation of motion for the nanoparticle,
\begin{equation}\label{eq:newton}
  m\ddot{\bm r}=\bm F_{\rm opt}(\bm r)+\bm F_{\rm drag}+\bm F_{\rm stoch}\,,
\end{equation}
where $m$ is the mass of the particle, which might include the added mass due to the fluid~\cite[Sec.~4.15]{newman:17}, $\bm r$ is the particle position, $\bm F_{\rm drag}$ the drag force of the particle moving through the fluid, and $\bm F_{\rm stoch}$ accounts for the stochastic fluid forces that are needed according to the fluctuation-dissipation theorem to counterbalance the drag forces~\cite{kubo:85}.  By successively computing the optical forces and updating the particle position according to Eq.~\eqref{eq:newton}, we obtain the nanoparticle trajectories.  Altogether, the theoretical model for \textsc{of2}i can be broken up into the following four steps.

\begin{enumerate}
\item Provide an expression for the incoming electromagnetic fields of the Laguerre-Gauss laser beam.

\item Solve Maxwell's equations in presence of the nanoparticle, using either an analytical or numerical approach.  This step provides us with the scattered electromagnetic fields.

\item Use the total fields, this is the sum of incoming and scattered fields, to compute the optical force acting on the nanoparticle at a given position.

\item Use Newton's equation of motion including optical and microfluidic forces to obtain the particle trajectory.
\end{enumerate}

\noindent In this work we will establish the methodology for this four-step model and discuss results of representative simulation setups.  In the future we plan to extend this model by tracing the scattered electromagnetic fields through the imaging system, which will allow us a most direct comparison with the experimental results.  For completeness, we here list the additional steps that will be needed to simulate the imaging system.

\begin{enumerate}
\setcounter{enumi}{4}

\item Propagate scattered electromagnetic fields through glass boundaries of microfluidic flow cell.

\item Simulate imaging of scattered electromagnetic fields, using for instance the approach of Richards and Wolf~\cite{richards:59,novotny:06,hohenester:20}.
\end{enumerate}

\noindent We start by discussing the electromanetic part of our simulation approach.  The power scattered by the nanoparticle is computed from the flow of scattered energy through the nanoparticle boundary~\cite{jackson:99}
\begin{equation}\label{eq:psca}
  P_{\rm sca}=\frac 12\oint_{\partial V}\mbox{Re}\left(\bm E_{\rm sca}^{\phantom*}\times
  \bm H_{\rm sca}^*\right)\cdot d\bm a\,,
\end{equation}
where $\partial V$ is the particle boundary with the infinitesimal boundary element $d\bm a$.  In deriving this expression we have assumed the usual time harmonic dependence $e^{-i\omega t}$ for the electromagnetic fields and have averaged over an oscillation cycle.  Eq.~\eqref{eq:psca} gives an estimate of how bright the nanoparticle appears in an imaging system, although a detailed analysis should additionally include the emission pattern of the scattered fields and the aforementioned deflection of these fields through lenses.  

Similarly, the transfer of momentum from the electromagnetic fields to the nanoparticle, this is the optical force, can be computed from the net flux of momentum carried by the electromagnetic fields through the nanoparticle boundary and by utilizing momentum conservation in the composite system formed by the nanoparticle and the electromagnetic fields.  This is, the inbalance of electromagnetic flux through the nanoparticle boundary provides us with the momentum transferred from the fields to the nanoparticle.  For time harmonic electromagnetic fields and by averaging over an oscillation cycle, we obtain under the assumption of quasi-stationarity, where the nanoparticle motion is negligible on the time scale of the field oscillations, the expression~\cite{marago:13,jones:15,gennerich:17,hohenester:20}
\begin{equation}\label{eq:fopt}
  \bm F_{\rm opt}=\frac 12\oint_{\partial V}\mbox{Re}\left[
  \overset\leftrightarrow{\theta}-\frac 12\openone\mbox{tr}\big(\overset\leftrightarrow{\theta}\big)
  \right]\cdot d\bm a\,.
\end{equation}
The term in brackets is Maxwell's stress tensor accounting for the momentum density flow of the electromagnetic fields, with~\cite{jackson:99}
\begin{equation}\label{eq:stress}
  \theta_{ij}=\varepsilon E_i^{\phantom*}E_j^*+\mu H_i^{\phantom*}H_j^*\,,
\end{equation}
where $\varepsilon$ and $\mu$ are the permittivity and permeability of the embedding background medium, respectively.  Eqs.~\eqref{eq:psca} and \eqref{eq:fopt} are the central expressions for the electromagnetic part of our theoretical modeling, and can be evaluated once the electromagnetic fields are at hand.  Note that the expression for the optical force can be easily generalized to obtain optical torques acting on nanoparticles, which is of importance for non-spherical particle geometries~\cite{jones:15,gennerich:17,hohenester:20}.

For the trajectory part, we consider for the force on a small sphere moving with veclocity $\bm v$ through a viscous fluid the usual Stokes' drag valid for a creeping flow with a Reynolds number much smaller than one~\cite{note:stokes}
\begin{equation}
  \bm F_{\rm drag}=-6\pi\mu R_{\rm hyd}\big(\bm v-\bm v_{\rm fluid}\big)\,,
\end{equation}
where $\bm v_{\rm fluid}$ is the velocity of the fluid and $\mu$ the dynamic viscosity.  In this work we set for simplicity $R_{\rm hyd}$ to the radius of the sphere, but in general this hydrodynamic radius might differ from the radius entering the optical calculations~\cite{wyatt:14}.  We will address this point in future work.  

For sufficiently large spheres, say for diameters above 10 nm, the momentum relaxation time is so short that we can approximately set $\dot{\bm v}\approx 0$~\cite{neuman:08}.  Also the stochastic forces don't play a decisive role for larger spheres, as will be discussed in Sec.~\ref{sec:stochatsic}.  The nanosphere's velocity $\bm v$ is then obtained from the condition that the optical force is balanced by the drag force, and we get 
\begin{equation}\label{eq:vsteady}
  \bm v(\bm r)=\bm v_{\rm fluid}+\frac{{\bm F}_{\rm opt}(\bm r)}{6\pi\eta R_{\rm hyd}}\,.
\end{equation}
We emphasize that our model contains no free parameters, and all laser, fluid, and nanoparticle parameters can be inferred in principle from experiment.

\subsection{Mie theory}

\begin{figure}[t]
\includegraphics[width=\columnwidth]{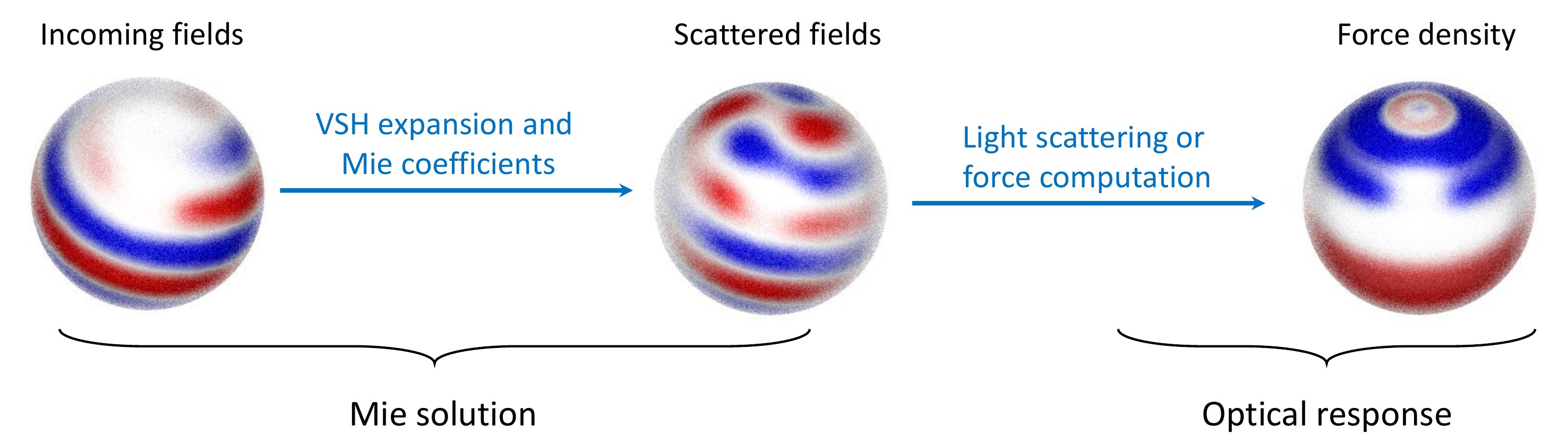}
\caption{Schematics of optical simulation approach.  The incoming fields of the vortex laser are expanded in vector spherical harmonics (\textsc{vsh}), and are used together with the Mie coefficients to compute the scattered electromagnetic fields.  Once the incoming and scattered fields are at hand, we can compute optical response properties such as the scattered light or the optical forces.  In the right panel we show the $z$-component of the force density, this is the integrand of Eq.~\eqref{eq:fopt}, on the sphere boundary.}
\label{fig:scheme}
\end{figure}

Mie theory provides an efficient and versatile method for solving Maxwell's equations for spherical nanoparticles~\cite{bohren:83}, as schematically depicted in Fig.~\ref{fig:scheme}.  The basic idea is to expand the electromagnetic fields in a complete basis with spherical symmetry.  The transverse fields can be expanded using~\cite{jackson:99,hohenester:20}
\begin{equation}\label{eq:basis}
  z_\ell(kr)\bm X_{\ell,m}(\theta,\phi)\,,\quad \nabla\times z_\ell(kr)\bm X_{\ell,m}(\theta,\phi)\,,
\end{equation}
where $z_\ell(kr)$ are spherical Bessel or Hankel functions, $k$ is the wavenumber of the medium, and $\bm X_{\ell,m}$ are the vector spherical harmonics.  The angular degree and order are denoted with $\ell$ and $m$, respectively.  The basis of Eq.~\eqref{eq:basis} has the advantage that field matching at the nanosphere boundary can be done easily and seperatly for each pair of $\ell$, $m$.  Unfortunately, Mie theory is often complicated by the fact that the definitions of the various functions are not unique and different choices have been adopted in the literature, such that it is often difficult to compare results.  We here largely follow the definitions given in~\cite{kiselev:14,jackson:99,hohenester:20}.  For the incoming fields we choose spherical Bessel functions, which become plane waves at large distances $kr\gg 1$.  The incoming electromagnetic fields can then be expanded via
\begin{eqnarray}\label{eq:mieinc}
  \bm E_{\rm inc}&=&\sum_{\ell,m}\left[b_{\ell,m}^{\rm inc}j_\ell\bm X_{\ell,m}+
  \frac ik a_{\ell,m}^{\rm inc}\nabla\times j_\ell\bm X_{\ell,m}\right]Z \nonumber\\
  \bm H_{\rm inc}&=&\sum_{\ell,m}\left[a_{\ell,m}^{\rm inc}j_\ell\bm X_{\ell,m}-
  \frac ik b_{\ell,m}^{\rm inc}\nabla\times j_\ell\bm X_{\ell,m}\right]\,,\quad
\end{eqnarray}
where $Z$ is the impedance and $a_{\ell,m}^{\rm inc}$, $b_{\ell,m}^{\rm inc}$ are the coefficients to be determined for specific incoming fields.  Similarly, for the scattered fields outside the nanoparticle we choose spherical Hankel functions, which become outgoing spherical waves at large distances,

\begin{eqnarray}\label{eq:miesca}
  \bm E_{\rm sca}&=&-\sum_{\ell,m}\left[b_{\ell,m}^{\rm sca}h_\ell^{(1)}\bm X_{\ell,m}+
  \frac ik a_{\ell,m}^{\rm sca}\nabla\times h_\ell^{(1)}\bm X_{\ell,m}\right]Z \nonumber\\
  \bm H_{\rm sca}&=&-\sum_{\ell,m}\left[a_{\ell,m}^{\rm sca}h_\ell^{(1)}\bm X_{\ell,m}-
  \frac ik b_{\ell,m}^{\rm sca}\nabla\times h_\ell^{(1)}\bm X_{\ell,m}\right]\,.\qquad
\end{eqnarray}
These scattered fields are uniquely determined upon knowledge of the coefficients $a_{\ell,m}^{\rm sca}$, $b_{\ell,m}^{\rm sca}$.  Additionally, we need the scattered electromagnetic fields inside the nanosphere, which are identical to Eq.~\eqref{eq:miesca}, however, with the replacement of the spherical Hankel by spherical Bessel functions that remain finite at the origin, and with different coefficients $c_{\ell,m}^{\rm sca}$, $d_{\ell,m}^{\rm sca}$.  Below we will discuss how the scattering coefficients can be obtained through field matching at the sphere boundary.

For the incoming fields we consider a weakly focused Laguerre-Gauss laser beam and employ the paraxial approximation~\cite{song:20}, which is well justified for our case of weak focusing.  The explicit expressions are given in Appendix~\ref{sec:LG}.  In~\cite{kiselev:14} the coefficients $a_{\ell,m}^{\rm inc}$, $b_{\ell,m}^{\rm inc}$ were computed by matching the incoming fields and the Mie expansion of Eq.~\eqref{eq:mieinc} in the far-field limit.  We here proceed somewhat differently and compute the coefficients using the field values on the sphere boundary~\cite[Eq.~(E.5)]{hohenester:20}
\begin{eqnarray}\label{eq:mieinc2}
  a_{\ell,m}^{\rm inc}j_\ell(kR) &=&-\frac{Z^{-1}k}{\sqrt{\ell(\ell+1)}}
  \oint Y_{\ell,m}^*\Big[\bm r\cdot \bm E_{\rm inc}(\bm r+\bm r_0)\Big]\,d\Omega\nonumber\\
  b_{\ell,m}^{\rm inc}j_\ell(kR) &=&\phantom-\frac{k}{\sqrt{\ell(\ell+1)}}
  \oint Y_{\ell,m}^*\Big[\bm r\cdot \bm H_{\rm inc}(\bm r+\bm r_0)\Big]\,d\Omega\,,\nonumber\\
\end{eqnarray}
where the integrals extend over the unit sphere, and $\bm r$ is a position determined by the unit sphere angles and located on the sphere with radius $R$.  In Mie theory, the coefficients have to be computed for a reference frame where the sphere center is in the origin.  As the incoming electromagnetic fields are defined in a reference frame where the focus is the origin, we have to translate $\bm r$ by the center position $\bm r_0$ of the nanosphere.  The computation of the integrals can be considerably accelerated by using an equidistant grid for the azimuthal coordinate and noting that the resulting integral can be computed using the fast Fourier transform~\cite{press:02}.  The remaining integral over the polar angle is computed by means of a Legendre-Gauss quadrature.  The implementation of Eq.~\eqref{eq:mieinc2} can be easily tested for an incoming plane wave through comparison with the resulting analytic expressions~\cite[Eq.~(10.53)]{jackson:99}.  

The computation of the scattered fields is particularly simple within Mie theory because each pair of angular degrees and orders $\ell$, $m$ can be handled separately.  Field matching is accomplished through the so-called Mie coefficients~\cite{bohren:83,hohenester:20}
\begin{eqnarray}\label{eq:miecoeffs}
  a_\ell&=&\frac{Z_2\psi_\ell(x_1)\psi_\ell'(x_2)-Z_1\psi_\ell'(x_1)\psi_\ell(x_2)}%
                 {Z_2\psi_\ell(x_1)\xi_\ell'( x_2)-Z_1\psi_\ell'(x_1)\xi_\ell(x_2)}\nonumber\\
  b_\ell&=&\frac{Z_2\psi_\ell'(x_1)\psi_\ell(x_2)-Z_1\psi_\ell(x_1)\psi_\ell'(x_2)}%
                 {Z_2\psi_\ell'(x_1)\xi_\ell(x_2)-Z_1\psi_\ell(x_1)\xi_\ell'(x_2)} \,,            
\end{eqnarray}
with $k_1$, $k_2$ being the wavenumbers of the medium inside and outside the nanosphere, respectively, and $Z_1$, $Z_2$ the corresponding impedances.  We have introduced the abbreviation $x=kR$ and the Riccati-Bessel functions $\psi_\ell(x)=xj_\ell(x)$, $\xi_\ell(x)=xh_\ell^{(1)}(x)$, where a prime indicates the derivative with respect to $x$.  With the Mie coefficients, the scattered and incoming fields can be related through
\begin{equation}\label{eq:abinc}
  a_{\ell,m}^{\rm sca}=a_\ell\,a_{\ell,m}^{\rm inc}\,,\quad
  b_{\ell,m}^{\rm sca}=b_\ell\,b_{\ell,m}^{\rm inc}\,.
\end{equation}
Thus, the entire solution of Maxwell's equations for spherical particles is embodied in the Mie coefficients of Eq.~\eqref{eq:miecoeffs}, where the matching of fields at the particle boundary has been explicitly worked out.  Mie theory can be also used to compute the optical forces from the incoming and scattering coefficients only.  We here follow the approach of \cite{gutierrez-cuevas:18} where analytic expressions are derived.  Appendix~\ref{sec:mie} gives the explicit formulas used in this work.

\subsection{Boundary element method}

We additionally performed simulations using a fully numerical Maxwell solver.  In this work these simulations are mainly used for testing purposes to check the proper implementation of our Mie theory.  However, in future work such an approach might be useful for the investigation of non-spherical or coupled particles.  We employ our home-made \textsc{nanobem} solver~\cite{hohenester.cpc:22} which is based on a boundary element method (\textsc{bem}) approach that can be easily adopted for the nanospheres under study.  Details of the approach and typical runtime examples are discussed in some length in~\cite{hohenester.cpc:22}.  In the present work we use the \texttt{optforce} function of the \texttt{galerkin.solution} class in order to directly compute the optical forces.  Results of our \textsc{bem} simulations will be presented in the next section.

\section{Results}\label{sec:results}

\begin{figure}[t]
\includegraphics[width=\columnwidth]{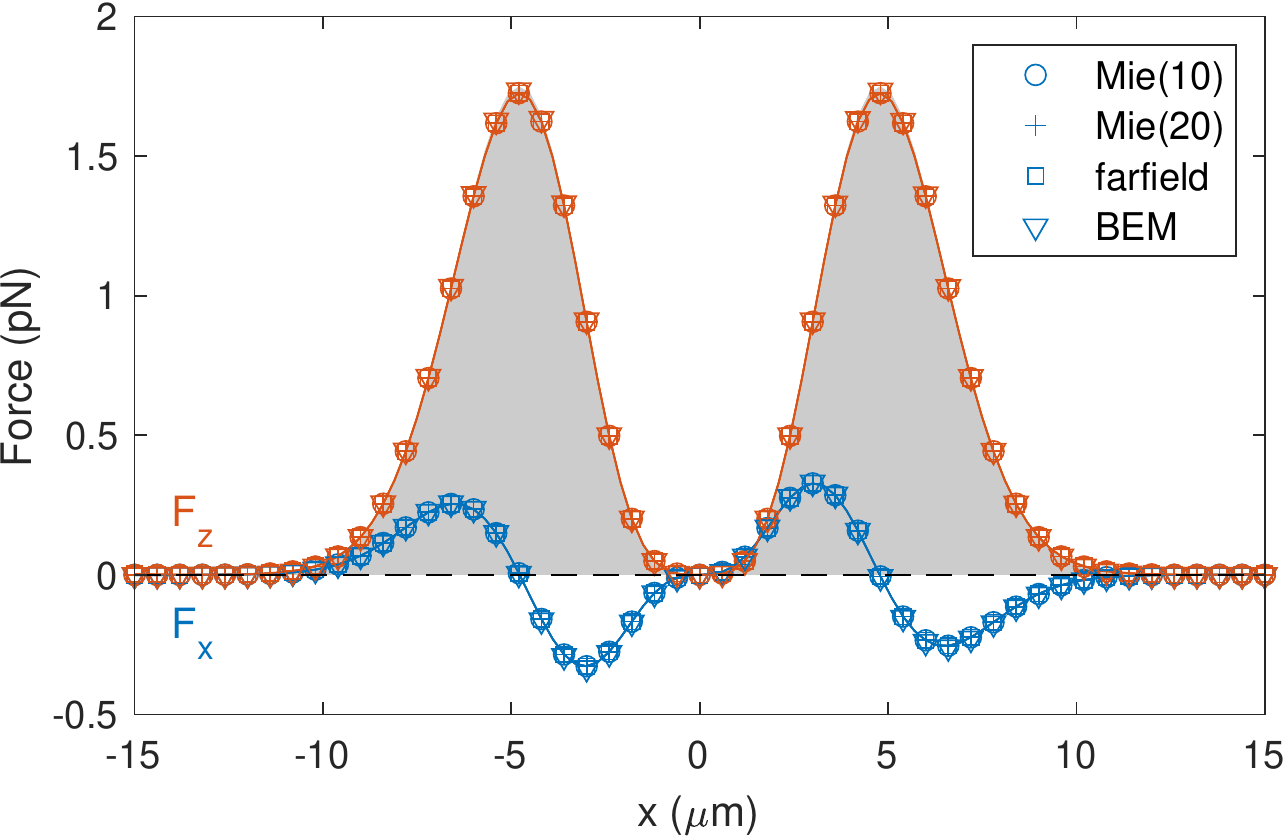}
\caption{Optical force $F_x$, $F_z$ in the focus plane $z=0$ and for a nanosphere with a diameter of 500 nm.  We compare results for different computation schemes.  \texttt{Mie}($\ell_{\rm max}$) report results for Mie theory with the cutoff number $\ell_{max}$ for the angular order and for the incoming fields computed within the paraxial approximation given in Appendix~\ref{sec:LG}.  \texttt{farfield} gives the results for the approach of~\cite{kiselev:14} where the fields are matched in the farfield, for details see text.  \texttt{BEM} reports results derived with our \textsc{nanobem} Maxwell solver based on a boundary element method approach.  For the sphere discretization we use 796 boundary elements, for details see~\cite{hohenester.cpc:22}.  The region shaded in gray reports the intensity of the Laguerre-Gauss laser beam in arbitrary units.  As apparent from the figure, the optical force in the propagation direction $F_z(x)$ directly follows the intensity profile.}
\label{fig:optforce02}
\end{figure}

Using the methodology developed in the previous section, we performed simulations with the same parameters as previously used in~\cite{simic:22}.  We consider a Laguerre-Gaussian beam with a topological charge of $m = 2$, a beam waist of $w_0=4.78$ $\mu$m for the fundamental Gaussian beam, a wavelength of $\lambda=532$ nm, and a power of 1.65 W.  For details of the incoming laser fields see Appendix~\ref{sec:LG}.  The fluid velocity is set to $v_{\rm fluid}=0.3$ mm/s and we use material parameters representative of water, namely a dynamic viscosity of $\eta=9.544 \times 10^{-4}$~Pa\,s and a refractive index of $n_b = 1.33$.  The refractive index of the nanospheres is set to $n=1.59$, a value representative for polystyrene, if not noted differently.

Figure~\ref{fig:optforce02} reports results for the optical force in the focus region.  The force $F_z$ in the longitudinal direction is largest at the intensity maxima of the vortex beam, see Fig.~\ref{fig:setup}.  There the sphere is pushed in the positive $z$ direction leading to the velocity enhancements to be discussed below.  The force $F_x$ in the transverse direction leads to trapping along $x$, and vanishes at the trapping positions $\pm w_0$,  where the intensity and $F_z$ is largest.  Additionally, there is an unstable equilibrium position at $x=0$ where no force is present because of the ring-like intensity profile of the vortex beam.  In the figure we compare different computation schemes, namely Mie theory with different cutoff numbers for the angular order, the determination of the incoming Mie coefficients using either Eq.~\eqref{eq:abinc} or the scheme presented in~\cite{kiselev:14}, and a fully numerical approach based on the boundary element method.  All schemes give indistinguishable results, thus demonstrating the accuracy and robustness of our approach.  

\begin{figure}[t]
\includegraphics[width=\columnwidth]{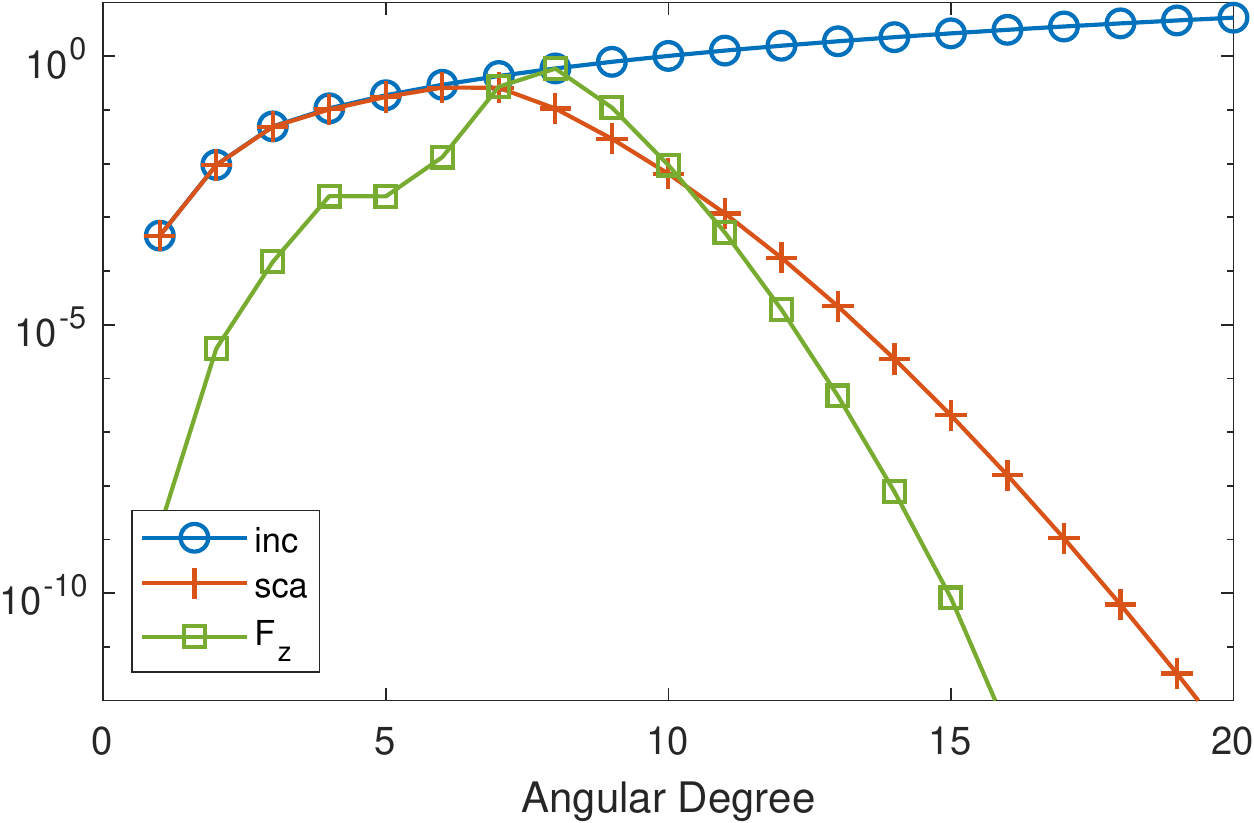}
\caption{Absolute values of incoming and scattering Mie coefficients, and of force $F_z$ as a function of angular degree $\ell$.  We consider a sphere with 1000 nm diameter at the trapping position in the focus plane.  For the incoming Mie coefficients we plot $\sum_m\left(\left|a_{\ell,m}^{\rm inc}\right|+\left|b_{\ell,m}^{\rm inc}\right|\right)$, with a similar expression for the scattering coefficients.  The contributions are scaled such that the sum of the scattering coefficients gives one. For the optical force, we report the increments $|F_z(\ell)-F_z(\ell-1)|$ for different degrees $\ell$.   The force contributions are scaled such that the sum gives one.}
\label{fig:optforce03}
\end{figure}

Figure~\ref{fig:optforce03} shows as function of the angular degree $\ell$ the absolute values of the incoming and scattered Mie coefficients for a nanosphere with 1000 nm diameter, which is trapped in the focus plane.  With increasing $\ell$ the incoming coefficients increase, whereas the Mie coefficients of Eq.~\eqref{eq:miecoeffs} decrease (not shown).  The scattering coefficients of Eq.~\eqref{eq:abinc} are the product of the incoming and Mie coefficients, which have a maximum at $\ell=6$ for the diameter under investigation, and then drop rapidly.  A similar behavior is observed for the optical force $F_z$, the explicit expressions are given in Appendix~\ref{sec:mie}.  In what follows, we choose a conservative cutoff number $\ell_{\rm max}=30$ for the angular degree, which provides a good compromise between fast simulations and highly accurate results.

\begin{figure*}
\centerline{\includegraphics[width=1.85\columnwidth]{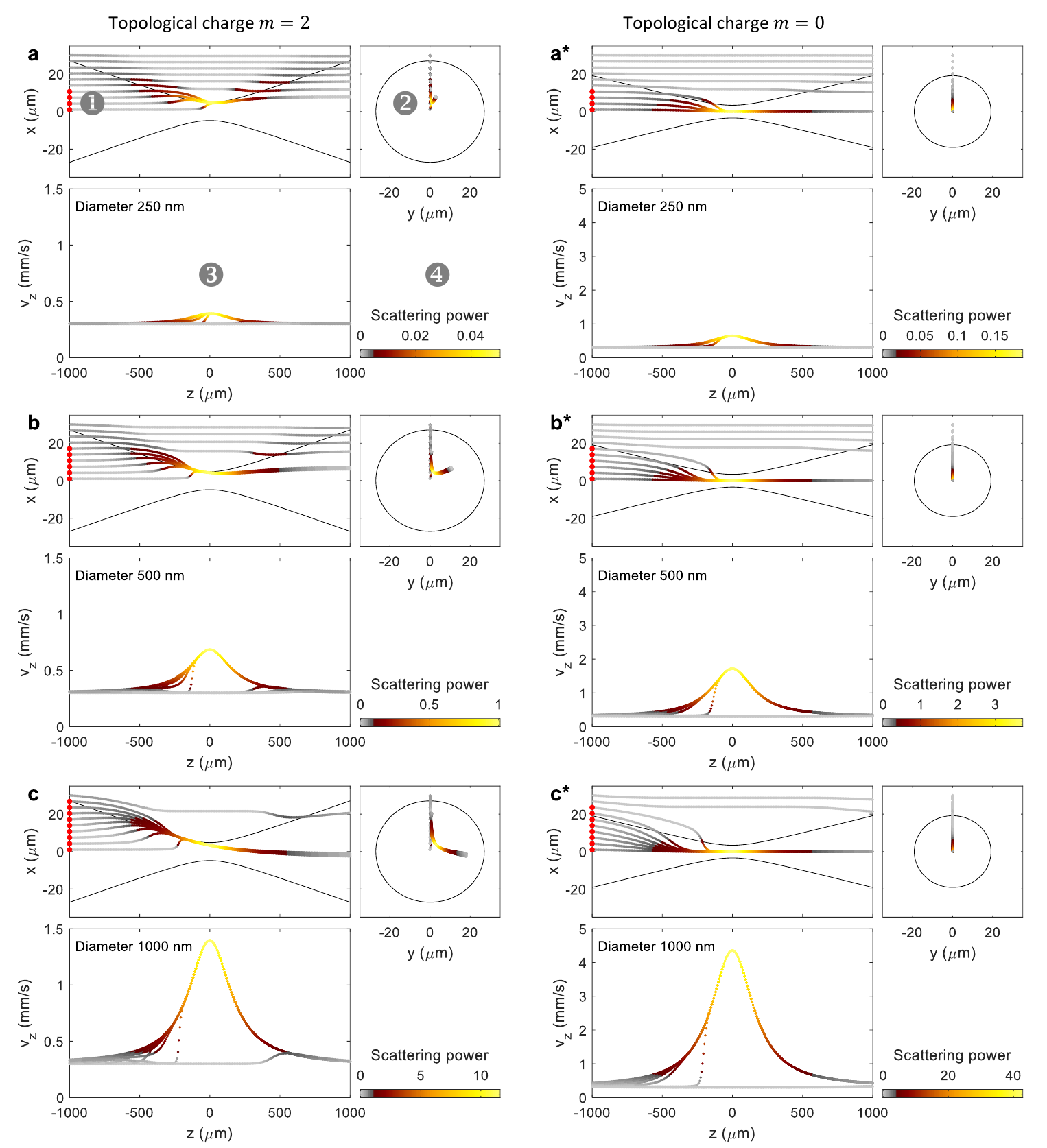}}
\caption{Trajectories and velocities for nanospheres with different diameters and for laser beams (a--c) with and (a*--c*) without an optical vortex.  In each panel, we report selected trajectories in the (1) $xz$ and (2) $xy$ plane, (3) the nanoparticle velocities as a function of propagation length $z$.  The colors of the line segments scale with the total scattering power of the spheres, given in arbitrary units with the colorbar reported in panel (4).  Trapped particles scatter more light and can be observed more easily.}
\label{fig:trajectory01}
\end{figure*}

Figure~\ref{fig:trajectory01} shows results for the nanosphere trajectories as obtained with the four-step model introduced in Sec.~\ref{sec:fourstep}.  We compare laser excitations (a--c) with and (a*--c*) without an optical vortex, as well as sphere diameters of (a) 250, (b) 500, and (c) 1000 nm.  Let us start by analyzing  the sub-figures of the various panels in slightly more detail.  In (1,2) we show selected trajectories.  Initially, the spheres are located at postions $(x,0,z_0)$ sufficiently far away from the focus ($z=0$) in a region where the optical forces are weak and can be neglected.  The nanoparticles are then transported through the fluid into regions of larger field strength, where some of them become trapped in the transverse directions.  The velocity changes of the trapped nanospheres in the laser propagation direction $z$ are shown in (3).  The color of the trajectories and velocities corresponds to the scattering power of Eq.~\eqref{eq:psca}, see (4) for the color code in arbitrary units.  It is apparent that trapped particles scatter more light and appear significantly brighter in an imaging system.  In the focus region, the scattered power of the trapped spheres with a diameter of 250 nm is at least three orders larger than that of the untrapped ones, and at least five orders for the larger spheres.  Additionally, only the trapped particles experience noticeable velocity changes.  The red dots in (1) indicate those particles which are trapped in the focus plane $z=0$.  As can be seen, some spheres become trapped after the focus plane.

When comparing the results for different sphere diameters in panels (a--c) of Fig.~\ref{fig:trajectory01}, we observe that with increasing diameter (i) more particles become trapped and (ii) experience larger velocity enhancements.  This can be attributed to the larger optical forces for larger nanoparticles.  We also find that (iii) the trajectories of all trapped particles are practically indistinguishable, and (iv) the deflection of the particles out of the $xz$-plane increases with increasing diameters [see panels (2)].  This is due to the orbital angular momentum transferred from the vortex laser to the nanospheres.  Finally, (v) also the scattering power increases with increasing diameter.  All these observations are supported by the experimental findings reported in~\cite{simic:22}, and suggest a dynamics where the nanospheres become first trapped in the transverse directions, and then propagate along the intensity maxima of the focused laser in presence of almost identical optic and fluidic forces through the focus region.  Note that in typical experiments the nanoparticles initially don't propagate in a single plane but are randomly distributed, correspondingly they are also randomly distributed in the focus region around the circular intensity maximum distribution of the vortex beam.  This leads to the aforementioned suppression of collissions and blockage in comparison to laser excitations with an intensity maximum on the optical axis.

To make this point more explicit, in panels (a*--c*) we report results for a Laguerre-Gauss excitation with zero topological charge, $m=0$, this is, for an excitation without an \textsc{oam}.  The trajectories are similar to the previous ones, with the exception of the larger velocity enhancements attributed to the higher field strengths of the focused laser without a vortex.  Additionally, we observe (2) that all particle trajectories are bound to the $xz$-plane because of the missing \textsc{oam}.  Owing to the laser intensity distribution that has a maximum at the $z$-axis for $m=0$, all trajectories are located on the $z$-axis around the focus regions, thus leading to particle collisions and blockage.

In what follows, we investigate the ability of \textsc{of2}i to infer from the observed velocity changes the size and material composition of the nanospheres.  We here only discuss the impact of these parameters and leave the problem of how to solve the inverse problem, namely the determination of size, material, and possibly geometry, to future work.

\subsection{Refractive index of nanospheres}

\begin{figure}
\centerline{\includegraphics[width=\columnwidth]{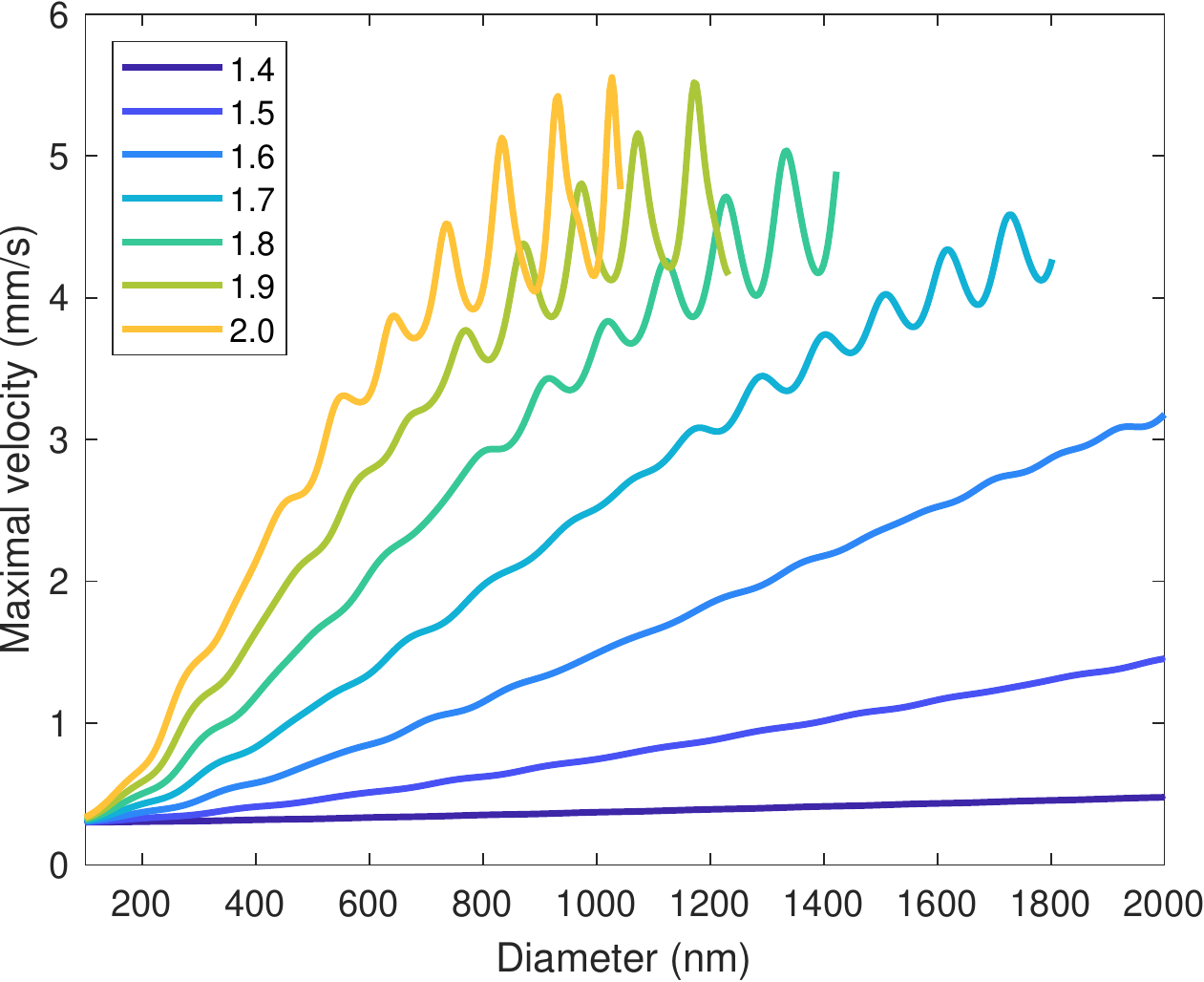}}
\caption{Maximal velocity in the focus region for nanospheres with different diameters and refractive indices (see inset).  For larger refractive indices the velocity increases non-monotonically because of Mie resonances supported by the spheres.}\label{fig:velfocus}
\end{figure}

Figure~\ref{fig:velfocus} shows the maximal velocity in the focus region for dielectric nanospheres with different diameters and refractive indices (see inset).  In all simulations we use water with an refractive index of $n_b=1.33$ for the embedding medium.  For the smallest refractive indices of the nanospheres, say for $n\le 1.6$, the maximal velocity increases monotonically with increasing diameter, at least for the sphere sizes under investigation.  In this regime it is thus possible to directly correlate the observed velocity enhancement with the particle diameter, as we have previously done in~\cite{simic:22}.  Things somewhat change for larger nanospheres where the optical response is governed by Mie resonances supported by the spherical nanoparticles.  Correspondingly, beyond a certain cutoff diameter the maximal velocity no longer simply increases with increasing diameter, but exhibits a more complicated resonance behavior.

\begin{figure}[t]
\centerline{\includegraphics[width=\columnwidth]{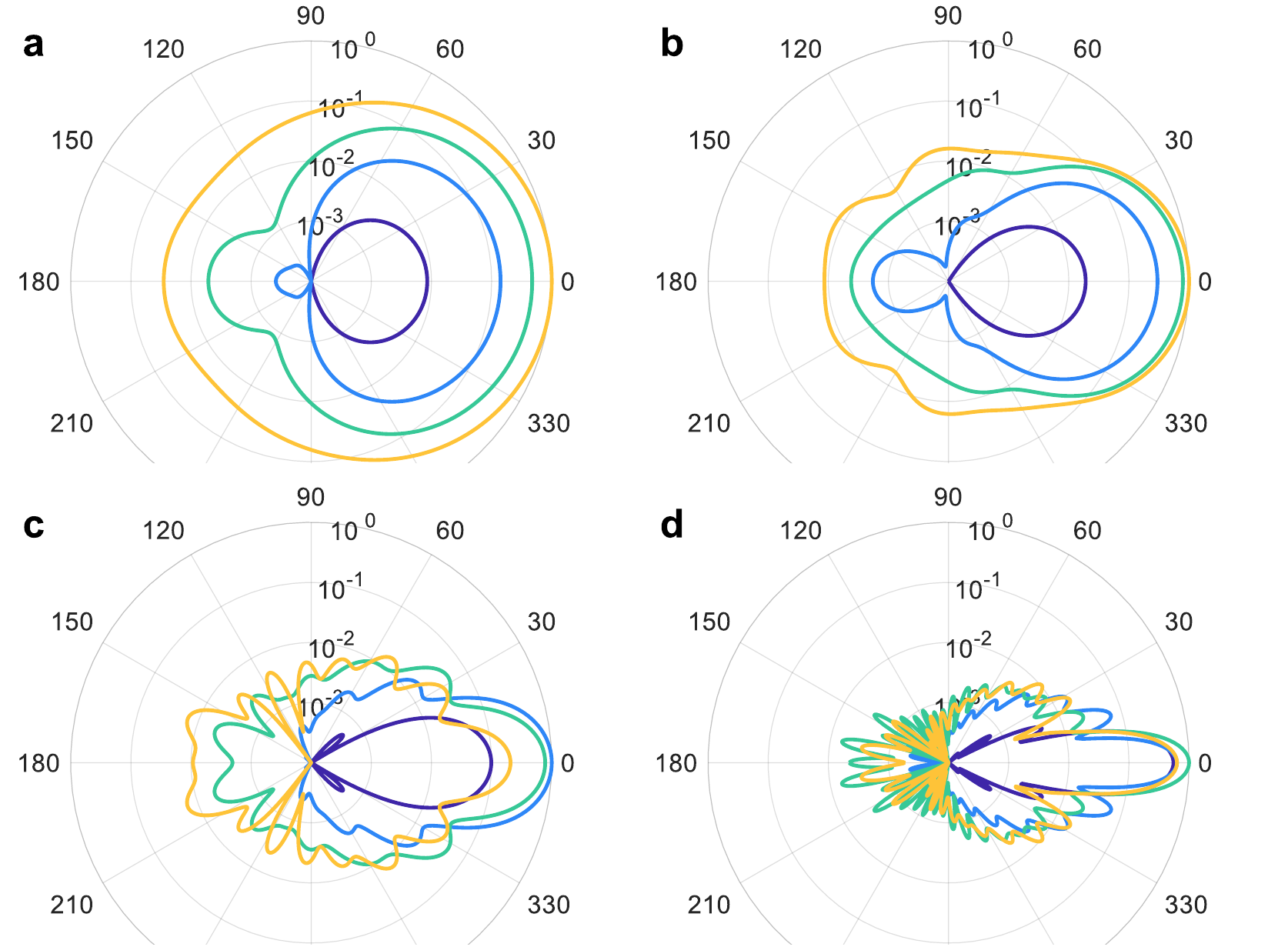}}
\caption{Normalized emission pattern for nanospheres with diameters of (a) 250, (b) 500, (c) 1000, and (d) 2000 nm.  We use a logarithmic scale in the radial direction and refractive indices of 1.4, 1.6, 1.8, 2.0, with the same color code as in Fig.~\ref{fig:velfocus}.  All plots are scaled to the respective maxima of the emission patterns.  In all cases the nanospheres are located in the focus plane at the trapping position around the intensity maxima of the vortex laser.}\label{fig:emission}
\end{figure}

For nanoparticles with larger refractive indices and/or larger particles in the micrometre range, in general it thus might be useful to analyze more carefully the light scattered off the nanoparticles.  In Fig.~\ref{fig:emission} we show the emission pattern of nanospheres with different diameters and refractive indices.  With increasing diameter the emission pattern sharpens into the forward direction (note that in the plots we use a logarithmic scale), however, at the same time the emission into other directions becomes strongly structured and provides detailed information about the nanosphere properties.  Using Fraunhofer diffraction and Mie scattering approaches, the characterization of particle sizes upon knowledge of the refractive indices of the nanoparticle and the embedding medium is a well established technique~\cite{boer:87}.  A more refined modeling of imaging within \textsc{of2}i would be needed (steps 5 and 6) to address the question whether the viable nanoparticle parameters can be uniquely extracted using this additional information.

\subsection{Active volume}

When inferring the particle number distribution from \textsc{of2}i measurements, we have to account for the fact that larger particles become trapped more easily than smaller ones, owing to the increase of optical forces with increasing particle size.  See for instance the red dots in panels (1) of Fig.~\ref{fig:trajectory01} for those particles which are trapped in the focus plane.  Recall that in our simulations we start with an initial position $(x,0,z_0)$ for the particles, where the propagation distance $z_0$ is located in a region where the optical forces are negligible (we use $z_0=-1$~mm).  Subsequently, the particles are transported by the fluid into regions of larger field intensities, where they become trapped and experience the velocity changes previously discussed. 

\begin{figure}[t]
\centerline{\includegraphics[width=0.9\columnwidth]{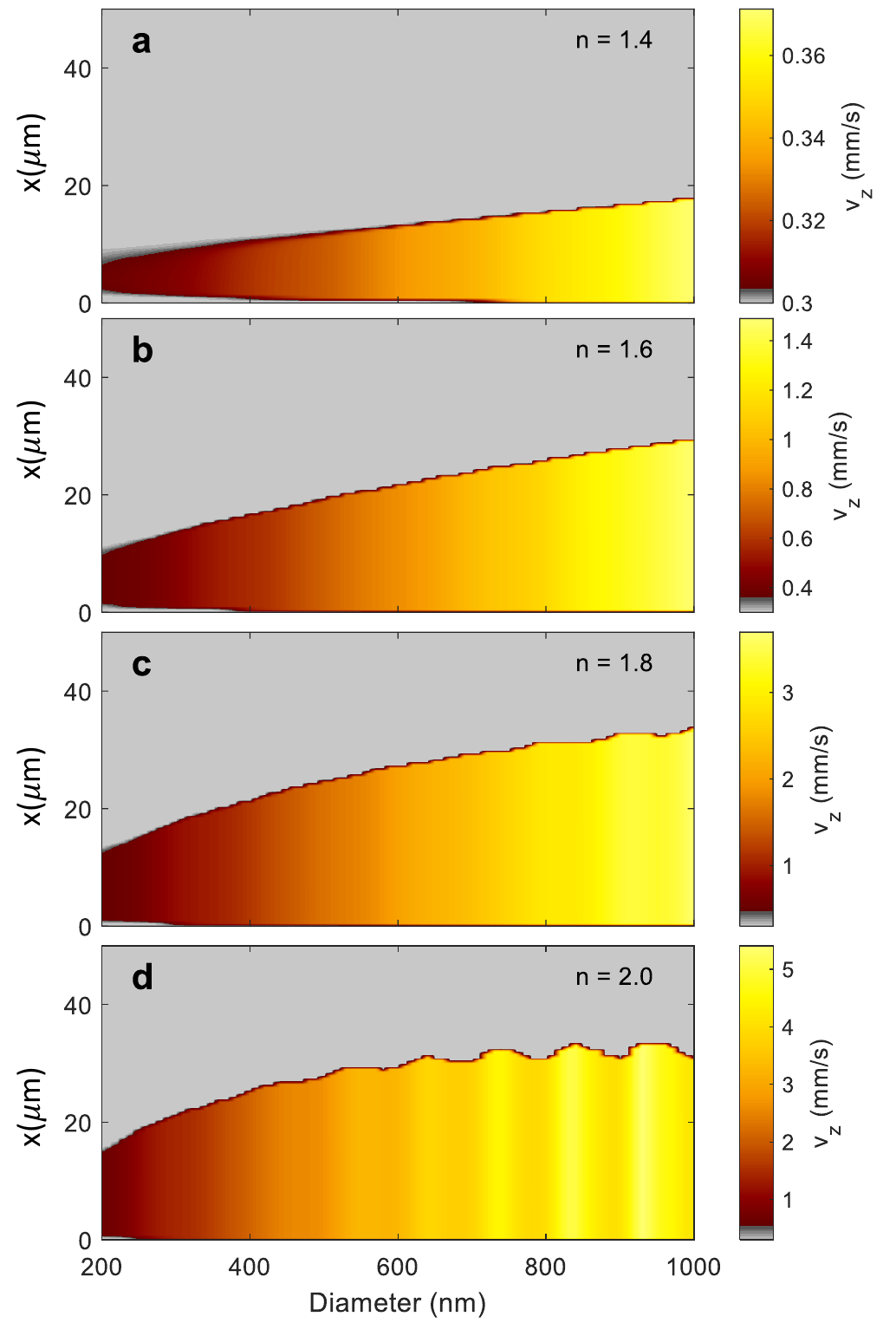}}
\caption{Velocity in focus region for different transverse starting positions $x$ and diameters, as well as for different refractive indices.  In all simulations the particles start at $(x,0,z_0)$ in a region where the optical forces are negligible.  Particles become either trapped or not (gray region), where all trapped particles are transported with the same velocity through the focus region.  With increasing sphere diameter more particles become trapped, owing to the increase of optical forces for the larger particles.}\label{fig:xcut}
\end{figure}

In Figure~\ref{fig:xcut} we show the velocities in the focus plane as a function of transverse starting position $x$ and sphere diameter, and for different refractive indices.  We observe that particles become either trapped or not, and for a given diameter and refractive index all trapped particles are transported with the same velocity through the focus plane.  This observation agrees with the velocity curves shown in panel (3) of Fig.~\ref{fig:trajectory01}.  When measuring particle size distributions one has to account for the different cutoff parameters for trapping $x_{\rm cut}(R,n)$, which depend on particle radius $R$ and refractive index $n$.  For starting position $x\le x_{\rm cut}$ particles are trapped in the focus plane, for $x> x_{\rm cut}$ the optical forces are too weak for trapping.  As previously discussed in~\cite{simic:22}, one can define an active volume
\begin{equation}
  V_{\rm active}(R,n)=\Big[\pi x_{\rm cut}^2(R,n)\Big]v_{\rm fluid}t_{\rm meas}\,,
\end{equation}
where the term in brackets is the cross section in the transverse direction, and $v_{\rm fluid}t_{\rm meas}$ is the size of the sampling volume along the propagation direction in the measurememt time $t_{\rm meas}$.  The active volume corrects for the fact that larger particles are trapped more easily and are observed more frequently in comparison to smaller particles.  For $N_{\rm meas}$ velocity counts within $t_{\rm meas}$, the particle density is then proportional to $\nicefrac{N_{\rm meas}}{V_{\rm active}}$.  

\subsection{Stochastic forces}\label{sec:stochatsic}

\begin{figure}[t]
\centerline{\includegraphics[width=1.05\columnwidth]{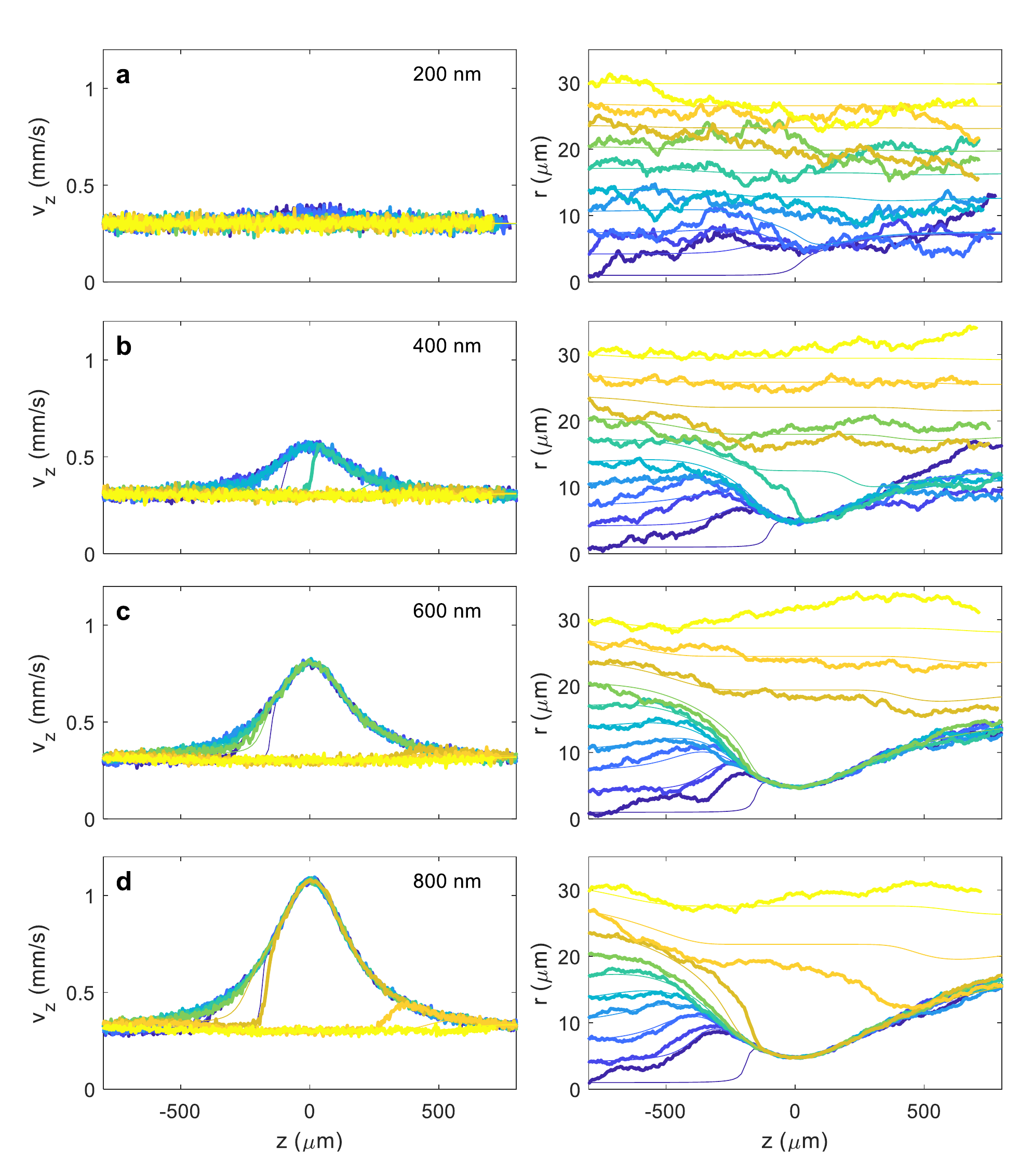}}
\caption{Velocities (left) and trajectories (right) as a function of propagation distance, with (thick lines) and without (thin lines) consideration of Brownian motion and for different sphere diameters (see inset).  We use different colors for the different starting positions of the spheres.  For the Brownian motion the velocity $v=\nicefrac{\Delta z}{\Delta t}$ is defined as the ratio between the propagation distance $\Delta z$ travelled by a particle in a time interval $\Delta t=0.01$ s and the time interval $\Delta t$.  $r=\sqrt{x^2+y^2}$ is the transverse distance. For the smallest diameter shown in panel (a) the stochastic forces are of equal strength than the optical forces.  For the larger diameters shown in panels (b--d) only the positions where the particles become trapped are somewhat altered by the Brownian motion.  Once they are trapped, they essentially follow the trajectories previously discussed for simulations without stochastic forces.}\label{fig:brownian}
\end{figure}

We finally comment on the influence of stochastic forces and Brownian motion, which are known to have an important impact for optical tweezers and related experiments.  The necessity for considering such forces was first noticed in the groundbreaking paper of Albert Einstein on Brownian motion~\cite{einstein:05}.  In our implementation of a stochastic force term we closely follow Ref.~\cite{bui:17}.  We first compute the drift velocity $\bm v$ using Eq.~\eqref{eq:vsteady} and then update the position according to~\cite[Eq.~(18)]{bui:17}
\begin{equation}
  \bm r(t+\Delta t)\approx \bm r(t)+\bm v\Delta t+\left(\frac{k_BT\delta t}{3\pi\eta R}\right)^{\frac 12}\bm W\,,
\end{equation}
where $\Delta t$ is the computational timestep, $k_B$ is Boltzmann's constant, $T$ is the temperature, $R$ is the sphere radius, and $W_x$, $W_y$, $W_Z$ are normally distributed random numbers with a variance equal to one, as obtained for instance by the \textsc{matlab} function \texttt{randn}.  The time step $\Delta t$ has to be chosen sufficiently small such that the optical forces at $\bm r(t)$ and $\bm r(t+\Delta t)$ do not differ significantly.  In all our simulations we used a value of $\Delta t=1$ ms and a temperature of $T=293$ K.

Fig.~\ref{fig:brownian} shows results for simulations including stochastic forces.  Let us first concentrate on the results for spheres with a sufficiently large diameter, say panels (b--d).  In contrast to simulations without stochastic forces (thin lines), the velocity curves exhibit fluctuations that decrease with increasing diameter, and the motion in the transverse direction is altered in regions of weak optical forces.  Once particles are trapped, they follow along the intensity maxima of the laser along trajectories that are very similar to the ones we have previously discussed.  As in \textsc{of2}i experiments predominantly the trapped particles can be observed, stochastic forces typically have no crucial impact on the observed particle trajectories.  Things are somewhat different for the smallest spheres, where the stochastic forces are of equal strength than the optical forces, and trapping can only be observed close to the focus region.  Such behavior is not found in experiment where spheres with a diameter of 200 nm are clearly trapped.  We attribute this diagreement to our simplified choice of the hydrodynamic radius in Eq.~\eqref{eq:vsteady}, and will analyze this point in more detail elsewhere.

\section{Summary and Outlook}\label{sec:summary}

To summarize, we have presented a four-step model for the theoretical description of \textsc{of2}i, which accounts for the nanoparticle propagation in a microfluidic channel in presence of laser excitation.  The approach is currently based on Mie theory but can be extended with moderate computational overhead to full Maxwell solvers, using for instance the boundary element method, in order to simulate non-spherical or coupled particles.  We have investigated the influence of particle size, refractive index, and Brownian motion on the observed trajectories and velocity enhancements.  Quite generally, our results support the unique measurement capabilities of \textsc{of2}i for single-particle tracking with high throughput.  

\textsc{of2}i measurement results provide additonal information such as the emission pattern, which might be used in future work to extract further properties of the particles to be analyzed.  With this additional information we might overcome the difficulties regarding Mie resonances, in particular for particles with larger refractive indices, which currently lead to a problematic non-monotonic relation between sphere diameter and velocity enhancement.  It will be also interesting to see how our conclusions become modified for non-spherical particles or particles with no sharp interfaces.

From the experimental side, we plan to investigate shorter Rayleigh ranges where smaller particles can be trapped more easily, as well as different polarization states of the incoming laser.  For small particles the issue regarding geometric and hydrodynamic radius should be addressed with greater care.  We also expect that for absorbing particles heating effects and the resulting photophoretic forces must be taken into account.  This leaves us with a relatively large to-do list for the future.  However, the four-step model introduced in this work provides us with a solid and versatile machinery for future investigations.

\section*{Acknowledgements}

This work was supported in part by the Austrian Research Promotion Agency (FFG) through project AoDiSys 891714, the European Commission (EC) through the projects NanoPAT (H2020-NMBP-TO-IND-2018-2020, Grant Agreement number: 862583) and MOZART (HORIZON-CL4-2021-RESILIENCE-01, Grant Agreement Number: 101058450).  We thank the whole
nano-medicine workgroup at the Gottfried Schatz Research Center for their cooperation and most helpful discussions.

\begin{appendix}

\section{Fields of Laguerre-Gauss beam}\label{sec:LG}

The electromagnetic fields for a Laguerre-Gauss laser beam within the paraxial approximation are taken from~\cite{song:20} and are repeated in this Appendix for the sake of completeness.  Let $m$ be the topological charge of the vortex beam and $w_0$ the beam waist.  The radial index is set to $n=0$ throughout.  The wavenumber of the embedding medium is $k$.  We introduce the Rayleigh range
\begin{equation}
  z_R=\frac 12 kw_0^2
\end{equation}
and the $z$-dependent waist
\begin{equation}
  w(z)=w_0\sqrt{1+\zeta^2}\,,
\end{equation}
where $\zeta=\frac z{z_R}$.  We next define~\cite[Eq.~(3)]{song:20}
\begin{equation}
  u_0=\frac 1{1+i\zeta}\exp\left[-\left(\frac r{w_0}\right)^2\frac 1{1+i\zeta}\right]
\end{equation}
together with
\begin{equation}
  u_m=\left(\frac{\sqrt 2 r}{w(z)}\right)^m\exp\left[im\left(\phi-\tan^{-1}\zeta\right)\right]\,.
\end{equation}
The electric field is then given through~\cite[Eqs.~(35,37)]{song:20}
\begin{eqnarray}
  E_x &=& Au_0u_me^{ikz}\\
  E_z &=& \left(\frac{m(x+iy)}{kr^2}-\frac {ix}{iz-z_R}-\frac{4ix}{kw^2}\right)Au_0u_me^{ikz}\,.\nonumber
\end{eqnarray}
Here $A$ is the amplitude of the laser beam.  Similarly, the magnetic field reads~\cite[Eqs.~(39,49)]{song:20}
\begin{eqnarray}
  ZH_y &=& Au_0u_me^{ikz}\\
  ZH_z &=& \left(\frac{m(iy-x)}{kr^2}-\frac {iy}{iz-z_R}-\frac{4iy}{kw^2}\right)Au_0u_me^{ikz}\,.\nonumber
\end{eqnarray}

\section{Optical forces within Mie theory}\label{sec:mie}

In this Appendix we give the expressions for the optical forces in terms of Mie coefficients~\cite{gutierrez-cuevas:18}.  A few modifications arise due to the different notations adopted in this work.  We first introduce the abbreviations
\begin{eqnarray*}
  \Lambda^{(1)} &=& \frac 1{\ell+1}\sqrt{\frac{(\ell+m+2)(\ell+m+1)\ell(\ell+2)}%
  {(2\ell+1)(2\ell+3)}} \\
  \Lambda^{(2)} &=& \frac 1{\ell+1}\sqrt{\frac{(\ell-m+2)(\ell-m+1)\ell(\ell+2)}%
  {(2\ell+1)(2\ell+3)}} \\
  \Lambda^{(3)} &=& - \frac{\sqrt{(\ell+m+1)(\ell-m)}}{\ell(\ell+1)}
\end{eqnarray*}
as well as
\begin{eqnarray*}
  \Lambda_z^{(1)} &=& \frac 1{\ell+1}\sqrt{\frac{(\ell-m+1)(\ell+m+1)\ell(\ell+2)}%
  {(2\ell+1)(2\ell+3)}}\\
  \Lambda_z^{(2)} &=& \frac m{\ell(\ell+1)}\,.
\end{eqnarray*}  
The expressions given in~\cite[Eq.~(29a)]{gutierrez-cuevas:18} can then be written in the compact form
\begin{eqnarray*}
  f &=& \Lambda^{(1)}\left[2a_{\ell,m}^{\rm sca}a_{\ell+1,m+1}^{{\rm sca}\,*}+
                   		    a_{\ell,m}^{\rm inc}a_{\ell+1,m+1}^{{\rm sca}\,*}+
                            a_{\ell,m}^{\rm sca}a_{\ell+1,m+1}^{{\rm inc}\,*}\right] \\
    &+& \Lambda^{(1)}\left[2b_{\ell,m}^{\rm sca}b_{\ell+1,m+1}^{{\rm sca}\,*}+
                   		    b_{\ell,m}^{\rm inc}b_{\ell+1,m+1}^{{\rm sca}\,*}+
                            b_{\ell,m}^{\rm sca}b_{\ell+1,m+1}^{{\rm inc}\,*}\right] \\    
    &+& \Lambda^{(2)}\left[2a_{\ell+1,m-1}^{\rm sca}a_{\ell,m}^{{\rm sca}\,*}+
                   		    a_{\ell+1,m-1}^{\rm inc}a_{\ell,m}^{{\rm sca}\,*}+
                            a_{\ell+1,m-1}^{\rm sca}a_{\ell,m}^{{\rm inc}\,*}\right] \\
    &+& \Lambda^{(2)}\left[2b_{\ell+1,m-1}^{\rm sca}b_{\ell,m}^{{\rm sca}\,*}+
                   		    b_{\ell+1,m-1}^{\rm inc}b_{\ell,m}^{{\rm sca}\,*}+
                            b_{\ell+1,m-1}^{\rm sca}b_{\ell,m}^{{\rm inc}\,*}\right] \\                                
    &+& \Lambda^{(3)}\left[2a_{\ell,m}^{\rm sca}b_{\ell,m+1}^{{\rm sca}\,*}+
                   		    a_{\ell,m}^{\rm inc}b_{\ell,m+1}^{{\rm sca}\,*}+
                            a_{\ell,m}^{\rm sca}b_{\ell,m+1}^{{\rm inc}\,*}\right] \\
    &-& \Lambda^{(3)}\left[2b_{\ell,m}^{\rm sca}a_{\ell,m+1}^{{\rm sca}\,*}+
                   		    b_{\ell,m}^{\rm inc}a_{\ell,m+1}^{{\rm sca}\,*}+
                            b_{\ell,m}^{\rm sca}a_{\ell,m+1}^{{\rm inc}\,*}\right] \,.
\end{eqnarray*}
Similarly, we obtain ~\cite[Eq.~(29b)]{gutierrez-cuevas:18}
\begin{eqnarray*}
  f_z &=& \Lambda_z^{(1)}\left[2a_{\ell+1,m}^{\rm sca}a_{\ell,m}^{{\rm sca}\,*}+
                       		    a_{\ell+1,m}^{\rm inc}a_{\ell,m}^{{\rm sca}\,*}+
                                a_{\ell+1,m}^{\rm sca}a_{\ell,m}^{{\rm inc}\,*}\right] \\
      &+& \Lambda_z^{(1)}\left[2b_{\ell+1,m}^{\rm sca}b_{\ell,m}^{{\rm sca}\,*}+
                     		    b_{\ell+1,m}^{\rm inc}b_{\ell,m}^{{\rm sca}\,*}+
                                b_{\ell+1,m}^{\rm sca}b_{\ell,m}^{{\rm inc}\,*}\right] \\    
      &+& \Lambda_z^{(2)}\left[2b_{\ell,m}^{\rm sca}a_{\ell,m}^{{\rm sca}\,*}+
                       		    b_{\ell,m}^{\rm inc}a_{\ell,m}^{{\rm sca}\,*}+
                                b_{\ell,m}^{\rm sca}a_{\ell,m}^{{\rm inc}\,*}\right] \,.
\end{eqnarray*}
With these expression the optical force becomes
\begin{equation}
  \bm F_{\rm opt}=-\frac{\varepsilon_0}{2k^2}\left(
  \frac 12 \mbox{Im}[f]\,\hat{\bm x}-\frac 12\mbox{Re}[f]\,\hat{\bm y}+\mbox{Im}[f_z]\,\hat{\bm z}
  \right)\,.
\end{equation}
In setting up our Mie code with the above formulas we found it particularly useful to additionally perform \textsc{bem} simulations for excitations with a single $\ell$, $m$ term, and to compare the \textsc{bem} and Mie results.  With this comparison it is then relatively easy to check the proper implementation of the various contributions governing $\bm F_{\rm opt}$.

\end{appendix}

%\bibliography{of2i}

\begin{thebibliography}{34}
\bibitem[{not()}]{note:stokes}
\bibinfo{note}{For a flow in a circular pipe with a radius of $R$, the Reynolds
  number is defined as $\mbox{Re}=uR/\nu$, where $u$ is the mean velocity and
  $\nu$ the kinematic viscosity. Using values representative for our
  experiments, this is $R=0.5$ mm, $u=0.15\times 10^{-3}$ ms$^{-1}$, and
  $\nu=9.554\times 10^{-7}$ m$^2$s$^{-1}$, we obtain a Reynolds number of
  $\mbox{Re}\approx 0.08$.}
\end{thebibliography}

\end{document}